\documentclass[prl,10pt,twocolumn,showpacs,bibnotes,runinaddress]{revtex4}%
\usepackage{amsmath}
\usepackage{amsfonts}%
\usepackage{amssymb}%
\usepackage{graphicx}

\begin{document}
\title{Quantum Hall Effect under Rotation and Mass of the Laughlin Quasiparticles}
\author{Bo Zhao}
\affiliation{Department of Modern Physics, University of Science and Techlogy of China,
Hefei, Anhui 230027, China}
\author{Zeng-Bing Chen}
\email{zbchen@ustc.edu.cn}
\affiliation{Department of Modern Physics, University of Science and Techlogy of China,
Hefei, Anhui 230027, China}

\begin{abstract}
We consider the quantum Hall effect induced by magnetic field and rotation,
which can drive the Hall samples into the quantum Hall regime and induce
fractional excitations. Both the mass and the charge of the Laughlin
quasiparticles\ are predicted to be fractionally quantized. The observable
effects induced by rotation are discussed. Based on the usual Hall samples
under rotation, we propose an experimental setup for detecting\ the
macroscopic\ quantization phenomena\ and the fractional mass of the Laughlin quasiparticles.

\end{abstract}
\pacs{73.43.Cd}
\pacs{73.43.Cd}
\maketitle

In condensed matter, two-dimensional (2D) electron gases are of particular
interest due to the discovery of fractional quantum Hall (QH) effect by Tsui,
Stormer, and Gossard \cite{exp Tsui}. It is one of the spectacular macroscopic
quantization phenomens and has been investigated for two decades both
theoretically and experimentally \cite{Hall}. One of the exotic aspects is the
Laughlin quasiparticles of fractional charge $\frac{e}{l}$ ($l$ odd for
fundamental fractional\ filling fraction) \cite{Laughlin} which has been
demonstrated recently by several beautiful experiments \cite{fraction charge}.
Similar ideas have been explored in another strongly correlated system--dilute
cold neutral\ atoms in Bose-Einstein condensates (BECs) in recent years
\cite{theory Wilkin,zoller,Ho,Chen}. The rapidly rotating BECs, which can be
treated as\ 2D interacting atoms, will lie in the lowest landau level (LLL)
and the nondegenerate ground state is of the form of the famous\ Laughlin
state \cite{theory Wilkin,zoller}. In our previous paper \cite{Chen}, we
predicted some macroscopic quantization\ phenomena which are the
\textquotedblleft atomic twin of the electronic brother\textquotedblright\ on
the base of the similarity between these two strongly correlated
many-body\ systems. The quantized \textquotedblleft
mass-conductance\textquotedblright\ $\sigma^{(m)}$ and the quasiparticles of
fractional mass replace, respectively, the electron-conductance $\sigma^{(e)}$
and fractionally charged quasiparticles in the usual QH effect. But the
requirements for observing similar QH\ physics in BECs are demanding due to
the difficulty in the current BEC experiments\ to deposit a large enough
angular momentum to reach the QH regime.

Essentially, the phenomena predicted in Ref. \cite{Chen} in rapidly-rotating
stongly-correlated cold atoms are induced by the noninertial force on massive
particles in the rotating frame. Thus, these effects should also occur in
other strongly correlated many-body systems, even in the electron gases. The
detection of effects induced by rotation in magnetizable and metals has a
history of more than one hundred year \cite{Maxwell}. The famous experiment on
this respect was done by Barnett who measured the magnetic field induced by
rotation in $1915$ \cite{Barnett}. Recent experiments on the rotating effect
were to measure the mass of the current carriers (Cooper pairs) in
superconductors \cite{Zimmerman}. Recently, Fischer \textit{et al.
}\cite{Fischer}\textit{\ }have studied the rotation in QH samples using a
twofold $U(1)$ gauge invariance.

The charge of the current carriers (the Laughlin quasiparticles) in fractional
QH effect is fractionized as $\nu e$ for simple Hall droplet, as is now well
known. Now a fascinating question arises: Is the mass of the fractionally
charged quasiparticles in a fractional QH\ liquid also fractionized? To the
best of our knowledge, this question has not been investigated up to date. Its
answer is not as obvious as in a superconductor where the current carriers are
the Cooper pairs with mass $m_{CP}$ being duplation of electron's mass $m_{e}%
$, namely, $m_{CP}=2m_{e}$. In usual quantum Hall effect an excitation can be
induced by piercing a magnetic flux $\Phi_{e}=\frac{2\pi\hbar}{e}$. However,
there is no physical principle to determine the mass of the excitations.

Motivated by our previous work \cite{Chen}, in this paper we deal with such
kind of problems from a microscopic view. Namely, we are concerned with
QH\ effect under rotation. Of particular interest is the properties (e.g., the
mass) of the fractional excitations. By using the usual QH\ samples under
rotation, we propose an experiment to measure the fractional mass of the
Laughlin quasiparticles, which could be tested under current experimental conditions.

First, we consider a 2D rotating QH\ sample (see Fig. $1$) under a high
magnetic field $-B\hat{z}$, which is perpendicular to the 2D plane. Assume
that the sample is confined by a harmonic potential with trapping frequency
$\omega_{0}$ and then rotated, along the $\hat{z}$-direction, at an angular
frequency $\omega=\omega_{0}$. In the rotating frame the 2D electrons are
described by \cite{explanation}
\begin{equation}
H_{tot}=\sum\limits_{i}H_{i}+\sum\limits_{i<j}V(\mathbf{r}_{i}-\mathbf{r}%
_{j}).\label{ham}%
\end{equation}
Here $H_{i}=\frac{(\mathbf{p}_{i}-m\omega\widehat{z}\times\mathbf{r}%
_{i}\mathbf{+}e\mathbf{A}_{i})^{2}}{2m}$ is the single-body Hamiltonian;
$V(\mathbf{r}_{i}\mathbf{-r}_{j}\mathbf{)}$ describes the two-body
interaction, which in this case is the Columb interaction $\frac{e^{2}%
}{|\mathbf{r}_{i}-\mathbf{r}_{j}|}$ between electrons as is familiar in QH
effect. The Hamiltonian (\ref{ham}) can be naturally realized in a rotating
quantum-dot system under the same conditions specified above.

Now let us first consider the single-body Hamiltonian. As is done in our
previous paper \cite{Chen}, one can deal with the rotation and the magnetic
field in a uniform way, namely
\begin{equation}
q^{\ast}\mathbf{A}^{\ast}=m\omega\widehat{z}\times\mathbf{r+}e\mathbf{A,}%
\label{qa}%
\end{equation}
where $\mathbf{A}\mathbf{=-}\frac{1}{2}B\widehat{z}\times\mathbf{r}$ (with
$\mathbf{B}$ $=\nabla\times\mathbf{A}$ and $\mathbf{B}^{\ast}=\nabla
\times\mathbf{A}^{\ast}$) and $q^{\ast}$ is the effective charge. Then
single-body Hamiltonian is simplified to be
\begin{equation}
H=\frac{(\mathbf{p}-q^{\ast}\mathbf{A}^{\ast})^{2}}{2m}=\frac{[\mathbf{p}%
-(m\omega+\frac{eB}{2})\widehat{z}\times\mathbf{r}]^{2}}{2m},\label{h1}%
\end{equation}
which is completely analogous to the Hamiltonian with only effective magnetic
field $\mathbf{B}^{\ast}$. So when the magnetic field $B$ or/and the rotating
frequency $\omega$ is large enough, the system can be studied only in the
lowest landau level (LLL). In LLL the single-body states read $\psi_{l}%
(\eta)=N_{l}\eta^{l}\exp[-|\eta|^{2}/(4a_{0})^{2}]$ $(l=0,1,2,...)$ in terms
of the 2D complex coordinate $\eta=x+iy\equiv a_{0}\overline{\eta}$. Here
$N_{l}=[\pi l!(2a_{0})^{l+1}]^{-1/2}$ is the normalization constant and
$a_{0}=\sqrt{\hbar/(2m\omega+eB)}$ is the effective magnetic length.

We can also define the filling fraction as usual
\begin{equation}
\nu=\frac{2\pi n\hbar}{q^{\ast}B^{\ast}}=\frac{2\pi n\hbar}{2m\omega
+eB}.\label{ff}%
\end{equation}
In this case the many-body system will show some macroscopic quantum phenomena
due to the repulsive two-body interaction as in the electron or atom QH
effect. Its nondegenerate\ ground state is the Laughlin state
\begin{equation}
\Psi_{l}=\mathcal{N}_{l}\prod\limits_{i<j}(\overline{\eta}_{i}-\overline{\eta
}_{j})^{l}e^{-\frac{1}{4}\sum_{i}|\overline{\eta}_{i}|^{2}},\label{laughlin}%
\end{equation}
which was first put forward by Laughlin in a classic paper in 1983
\cite{Laughlin} to explain the fundamental filling fraction in the fractional
QH effect. Here $\mathcal{N}_{l}$ is an unimportant normalization constant,
and $l>0$ is an even number for bosons or an odd number for fermions. Then in
this state the filling fraction will be quantized as usual $\nu=\frac{1}{l}$.
The \textquotedblleft atomic Hall-conductance\textquotedblright\ in atomic QH
effect \cite{Chen} and the usual quantized Hall-conductance in electron
QH\ effect \cite{Hall} is naturally unified as
\begin{equation}
\frac{e^{2}}{\sigma^{(e)}2\pi\hbar}+\frac{m^{2}}{\sigma^{(m)}2\pi\hbar
}=l.\label{eml}%
\end{equation}
This means that neither of the two Hall conductances is quantized, but their
combination is quantized.

The equivalent electron-number current (the real current carriers are not the
electron but the quasiparticles, as will be discussed below) is
\begin{equation}
j_{x}=\frac{\nu}{2\pi\hbar}(eE_{y}+mg_{y}),\label{current}%
\end{equation}
where $E_{y}$ and $g_{y}$ are induced Hall electric field and
gravitational-like acceleration field, respectively. Equation (\ref{current})
is identical to the twofold $U(1)$ gauge invariant current obtained in Ref.
\cite{Fischer}. It means that when one inputs a particle current in the
sample, the current will induce not only the Hall electrostatic field but also
the gravitational-like acceleration field. The two induced fields are caused
by the two-body interaction. Then the equivalent particle\ current induced by
the change of magnetic field $B$ (piercing a magnetic flux $\Phi_{e}%
=\frac{2\pi\hbar}{e}$) or rotating velocity $\omega$ (piercing a vortex
$\Phi_{m}=\frac{2\pi\hbar}{m}$) can be described in a uniform way
\begin{equation}
j=\frac{dN}{dt}=\frac{\nu}{2\pi\hbar}\frac{d\Phi}{dt},\label{current2}%
\end{equation}
where $\Phi=m\oint\mathbf{\Omega}\times\mathbf{dr-}e\oint\mathbf{A\cdot
dr=}N_{v}2\pi\hbar$ is the total fluxes and $j$ is also the equivalent
electron-number current. So filling fraction naturally reads $\nu=\frac
{N}{N_{\nu}}$. This equation means that either piercing a magnetic flux
$\Phi_{e}=\frac{2\pi\hbar}{e}$ or piercing a vortex $\Phi_{m}=\frac{2\pi\hbar
}{m}$ in a disk or cylinder sample will induce a particle current as long as
the sample is in Hall regime.%
\begin{figure}
[ptb]
\begin{center}
\includegraphics[
height=1.6792in,
width=2.2151in
]%
{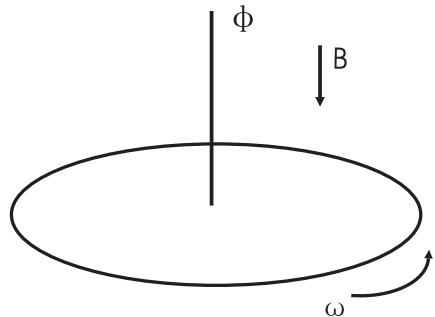}%
\caption{Schematic view of the rotating\ sample under high magnetic field
$\mathbf{B.} $}%
\label{view}%
\end{center}
\end{figure}

The exotic fractional excitations, which are the current carriers, are of
particular interest in electron QH effect \cite{Laughlin} and\ in atom QH
effect \cite{Chen}. What is the relation between these two kinds of
excitations? The answer to this question can be approached in the present
case. As usual, the fractional charge should read $Q^{\ast}=\frac{q^{\ast}}%
{l}$. In order to see clearly its meaning we add the external field $E^{\ast}%
$, which is induced by piercing flux $\Phi$ \cite{Chen} and reads
\begin{equation}
Q^{\ast}E^{\ast}=\frac{eE+mg}{l}=\frac{e}{l}E+\frac{m}{l}g.
\end{equation}
This relation means that the quasiparticles will respond to the electric field
as $\frac{e}{l}$ and acceleration field as $\frac{m}{l}$. This fact implies
that the quasiparticles are of both the fractional charge and the fractional
mass. Thus, piercing a magnetic flux $\Phi_{e}=\frac{2\pi\hbar}{e}$ and/or
piercing a vortex $\Phi_{m}=\frac{2\pi\hbar}{m}$ in the rotating\ Hall sample
will induce excitations from the unique vacuum state (i.e., the Laughlin
state) and lead to the same Laughlin quasiparticle state \cite{Laughlin,Chen}.
Consequently, the excitations possess simultaneously fractional charge and
fractional mass. This result is necessarily the property of the nondegenerate
ground state and does not depend on the origin of the ground state. So what we
got above should be correct even in usual nonrotating Hall sample under high
magnetic field $\mathbf{B}$, namely, the fractional excitations with charge
$\frac{e}{l}$ in electron QH effect possess the fractional mass $\frac{m}{l}$.

It seems that the electron-conductance will not be quantized as usual due to
the relation $I_{x}=\frac{\nu}{2\pi\hbar}(eV+m\Phi)$ ($\Phi$ is the induced
gravational-like potential \cite{Chen}, $I_{x}$ is the equivalent electron
current) if we add source and drain on the rotating QH sample. But this is not
the case. In fact, the quantum Hall\ voltage or the gravitational-like
potential are all caused by the two-body interaction, which is now the Columb
interaction. The gravational-like force is also\ induced by the two-body
interaction and behaves as another electrostatic field $eV^{\prime}\equiv
m\Phi$. What is actually measured is the total Hall voltage $V_{tot}%
=V+V^{\prime}$. Then the quantization relation in usual QH\ effect is formally
preserved
\begin{equation}
\sigma^{(em)}\equiv\frac{eI_{x}}{V_{tot}}=\nu\frac{e^{2}}{2\pi\hbar
}.\label{preserve}%
\end{equation}

Now let us consider how to measure the effects induced by rotation, in
particular, the fractional mass of the quasiparticles. This issue is currently
difficult for BECs for which the current experiments are far from the Hall
regime. We turn to other samples, such as the electron Hall sample, to find a
way to measure these effects for there have been some experiments to measure
the rotating effects and the mass of Cooper pairs in superconductors
\cite{Zimmerman}. In the following we will consider a special case and propose
a method to measure the physical effects induced by rotation and the
fractional mass of the Laughlin quasiparticles.%
\begin{figure}
[ptb]
\begin{center}
\includegraphics[
height=1.9285in,
width=2.2771in
]%
{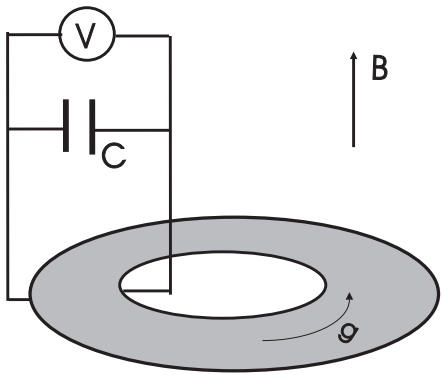}%
\caption{Schematic view of the Corbino sample to verify the fractional mass of
the Laughlin quasiparticles. Rotating the sample to induce $g$ will excite the
quasiparticles and induce a voltage in the capacitance. The voltage plateau
will demonstrate the fractional mass.}%
\label{Hall-m}%
\end{center}
\end{figure}

First we consider a usual Hall Corbino\ disk\ sample \cite{Hall}\ without
rotation. The Hamiltonian describing the system is
\begin{equation}
H_{tot}=\sum\limits_{i}H_{i}^{\prime}+\sum\limits_{i<j}V(\mathbf{r}%
_{i}-\mathbf{r}_{j}),
\end{equation}
where $H_{i}^{\prime}=\frac{(\mathbf{p}_{i}+e\mathbf{A}_{i})^{2}}{2m}$.
Laughlin's gedanken experiment is to pierce a magnetic flux $\Phi_{e}%
=\frac{2\pi\hbar}{e}$, which will induce a fractional charge transfer in the
Hall sample, as has already\ been demonstrated by experiments
\cite{Dolgopolov,Dolgopolovb,Jeanneret}. Based on the similar sample one can
also measure the mass of the fractionally charged quasiparticles which are the
current carriers in QH sample.

Let us consider the case of piercing a vortex $\Phi_{m}=\frac{2\pi\hbar}{m}$
in the sample by rotating the fractional Hall Corbino disk. Due to Faraday's
law \cite{Chen,Fischer}, this process will induce an acceleration field and
excite the electron gas. As described above it will induce a radial Hall mass
current \cite{Chen}
\begin{equation}
mj_{r}=\nu\frac{m^{2}}{2\pi\hbar}g_{\phi}.\label{qe}%
\end{equation}
Note that the current carriers are quasiparticles. Assuming $e^{\ast}$ and
$m^{\ast}$ are the charge and mass of the\ quasipaticles, respectively, one
can get the electric current $ej_{r}=\nu\frac{e^{\ast}}{m^{\ast}}\frac{m^{2}%
}{2\pi\hbar}g_{\phi}$ with $g_{\phi}=\frac{\overset{.}{\Phi}}{2\pi R}$. The
charge transferred through this process is
\begin{align}
Q &  =2\pi R\int ej_{r}dt=2\pi R\int dt\nu\frac{e^{\ast}}{m^{\ast}}\frac
{m^{2}}{2\pi\hbar}g_{\phi}\nonumber\\
&  =\nu\frac{e^{\ast}}{m^{\ast}}\frac{m^{2}}{2\pi\hbar}\int dt\overset{.}%
{\Phi}=\nu\frac{e^{\ast}}{m^{\ast}}\frac{m^{2}}{2\pi\hbar}\frac{2\pi\hbar}%
{m}\nonumber\\
&  =\nu e^{\ast}\frac{m}{m^{\ast}}.\label{q}%
\end{align}
According to the above results that $e^{\ast}=\frac{e}{l}$ and $m^{\ast}%
=\frac{m}{l}$ we will\ get $Q=\nu e$. This transferred charge is equal to
piercing a flux $\Phi=\frac{2\pi\hbar}{e}$. While the fractional charge of the
Laughlin quasiparticles\ has been demonstrated by several experiments through
measuring the shot noise in tunneling effect \cite{fraction charge}, the
fractional mass is unnoticed and thus, not determined up to date. So this
process opens a door to determine the fractional mass of the Laughlin quasiparticles.

For this purpose we propose a way to measure the proportion $\frac{e^{\ast}%
}{m^{\ast}}$ between the charge and the mass of the Laughlin quasiparticles to
demonstrate the fractional mass. The experimental device is depicted in Fig.
2. This device has been used in QH effect to test Laughlin's gedanken
experiment \cite{Jeanneret,Dolgopolovb}. The Corbino disk has reached the
fractional quantum Hall regime under high magnetic field $\mathbf{B}$
perpendicular to the 2D plane. Then rotating the sample to excite the
quasiparticles will induce charge to transfer from one edge to the other. The
charge transferred in this process is
\begin{equation}
Q=\frac{e^{\ast}}{m^{\ast}}\pi R^{2}\nu\frac{m^{2}}{2\pi\hbar}2\triangle
\omega,\label{qp}%
\end{equation}
where $R$ is the radius of the Corbino disk, $\triangle\omega$ is the rotation
angular velocity. This transferred charge can be detected by measuring the
voltage in capacitance, $V=\frac{Q}{C}$. Now the rotation only plays the role
of exciting the QH sample. Since the accuracy of the typical measurements in
QH\ effect is high, the rotation frequency does not need to be too large, say,
about $10^{3}\sim10^{4}s^{-1}$, is enough for current experimental conditions
\cite{Barnett,Zimmerman,Fischer} to observe the predicted effects. The methods
to observe similar phenomenon in QH effect by changing magnetic field in this
sample\ are familiar in the usual\ electron QH effect, where the\ quantized
conductance is measured. This result is also available to integer QH effect
where $\nu=\operatorname{integer}$ and the current carriers are electrons.

In summary, we have exploited the quantum Hall effect induced by magnetic
field and/or rotation. In this case both the magnetic field and the rotation
are available to reach the quantum Hall regime and can\ induce the fractional
excitations. Both the mass and charge of the quasiparticles, which are the
current carriers in the Hall sample,\ are fractionally quantized. The
observable effects induced by rotation have been discussed. An experimental
proof of\ the macroscopic\ quantization phenomena\ and the fractional mass of
the Laughlin quasiparticles\ has been proposed. We hope that\ our prediction
can be verified in the near future by using the usual quantum Hall samples
under rotation.

We thank Professor Yong-De Zhang for helpful discussion and U. P. Fischer for
bringing Ref. \cite{Fischer} into our attention. This work was supported by
the National Natural Science Foundation of China, the Chinese Academy of
Sciences and the Natural Fundamental Research Program (under Grant No. 2001CB309300).


\begin{thebibliography}{99}                                                                                               %
\bibitem {exp Tsui}D. C. Tsui, H. L. Stormer, and A. C. Gossard, Phys. Rev.
Lett. \textbf{48}, 1559 (1982).

\bibitem {Hall}R. E. Prange and S. M. Girvin, \textit{The} \textit{Quantum
Hall Effect }(Springer-Verlag, New York, 1990).

\bibitem {theory Wilkin}N. K. Wilkin, J. M. F. Gunn, and R. A. Smith, Phys.
Rev. Lett. \textbf{80}, 2265 (1998); N. R. Cooper and N. K. Wilkin, Phys. Rev.
B \textbf{60}, R16279 (1999); N. K. Wilkin and J. M. F. Gunn, Phys. Rev. Lett.
\textbf{84}, 6 (2000); N. R. Cooper , N. K. Wilkin, and J. M. F. Gunn, Phys.
Rev. Lett. \textbf{87}, 120405 (2001).

\bibitem {zoller}B. Paredes, P. Fedichev, J. I. Cirac, and P. Zoller, Phys.
Rev. Lett. \textbf{87}, 010402 (2001); B. Paredes, P. Zoller, and J. I. Cirac,
Phys.\ Rev. A \textbf{66}, 033609 (2002).

\bibitem {Ho}T.-L. Ho, Phys. Rev. Lett. 87, 060403 (2001); E. J. Muller and
T.-L. Ho, Phys. Rev. Lett. \textbf{88}, 180403 (2002); T.-L. Ho and E. J.
Muller, Phys. Rev. Lett. \textbf{89}, 050401 (2002).

\bibitem {Chen}Z.-B. Chen, B. Zhao, and Y.-D. Zhang, cond-mat/0211187.

\bibitem {Maxwell}J. C. Maxwell, \textit{A Treatise on Electricity and
Magnetism }( Dover, New York, 1954 ).

\bibitem {Barnett}S. J. Barnett, Phys. Rev. \textbf{6}, 239 (1916).

\bibitem {Zimmerman}A. F. Hildebrandt, Phys. Rev. Lett. 12, 190 (1964); J. E.
Zimmerman and J. E. Mercereau, Phys. Rev. Lett. \textbf{14}, 887 (1965); J.
Tate, S. B. Felch, and B. Cabrera, Phys. Rev. B \textbf{42}, 7885 (1990).

\bibitem {Fischer}U. R. Fischer and Schopohl, Europhys. Lett. \textbf{54}, 502
(2001); U. R. Fischer, C. H\"{a}ussler, J. Oppenl\"{a}nder, and N. Schopohl,
Phys. Rev B \textbf{64}, 214509 (2001).

\bibitem {explanation}This Hamiltonian also describles a rotating Hall sample
except for a centrifugal term which can be treated as an external $U(1)$
field, as is done by Fischer \textit{et al}. in \cite{Fischer}.

\bibitem {Laughlin}R. B. Laughlin, Phys. Rev. Lett. \textbf{50}, 1395 (1983).

\bibitem {Laughlinb}R. B. Laughlin, Phys. Rev. B \textbf{23}, 5632 (1981).

\bibitem {Dolgopolov}V. T. Dolgopolov, A. A. Shashkin, N. B. Zhitenev, S. I.
Dorozhkin, and K. von Klitzing, Phys. Rev. B \textbf{46}, 12560 (1992); V. T.
Dolgopolov, A A. Shashkin, G. V. Kravchenko, S. I. Dorozhkin, and K. von
Klitzing, Phys. Rev. B \textbf{48}, 8480 (1993).

\bibitem {Dolgopolovb}V. T. Dolgopolov and A. A. Shashkin, Phys. Rev. Lett.
\textbf{86}, 5566 (2001).

\bibitem {Jeanneret}B. Jeanneret, B. D. Hall, H.-J. B\"{u}hlmann, R.
Houdr\'{e}, and M. Ilegems, Phys. Rev. B \textbf{51}, 9572 (1995).

\bibitem {fraction charge}R. de-Picciotto, M. Reznikov, M. Heiblum, V.
Umansky, G. Bunin \& D. Mahalu, Nature (London) \textbf{389}, 162 (1997); L.
Saminadayar, D. C. Glarrli, Y. Jin and B. Erienne, Phys. Rev. Lett.
\textbf{79}, 2526 (1997); M. Reznlkov, R. de-Picciotto, T. G. Griffiths, M.
Heiblum and V. Umansky, Nature (London) \textbf{399}, 238 (1999).
\end{thebibliography}
\end{document}